\documentclass[a4paper,11pt]{article}

\pdfoutput=1 
\usepackage{jcappub}

\usepackage{graphicx}
\usepackage{longtable}
\usepackage{float}
\usepackage{dcolumn}
\usepackage{bm}
\usepackage{appendix}
\usepackage{multirow}
\usepackage{color}
\usepackage[utf8]{inputenc}
\usepackage{footmisc}
\usepackage[normalem]{ulem}

\newcommand{\mytilde}{\raise.17ex\hbox{$\scriptstyle\mathtt{\sim}$}}
\newcommand{\barr}{\begin{eqnarray}}
\newcommand{\earr}{\end{eqnarray}}
\newcommand{\bea}{\begin{eqnarray*}}
\newcommand{\eea}{\end{eqnarray*}}
\newcommand{\beq}{\begin{equation}}
\newcommand{\eeq}{\end{equation}}

\renewcommand{\bf}{\rm}
\newcommand{\MB}{\textcolor{black}}

\title{Dark sector interactions in light of weak lensing data}

\author[a,b]{M. Benetti}

\author[c]{P. T. Z. Seidel}

\author[d]{C. Pigozzo}

\author[e]{I. P. R. Baranov}

\author[d,f]{S. Carneiro}\emailAdd{saulo.carneiro.ufba@gmail.com}

\author[c]{J. C. Fabris}

\affiliation[a]{Scuola Superiore Meridionale, Largo S. Marcellino 10, I-80138 Napoli, Italy}

\affiliation[b]{Istituto Nazionale di Fisica Nucleare (INFN), sez. di Napoli, Via Cinthia 9, I-80126 Napoli, Italy}

\affiliation[c]{PPGCosmo, CCE, Universidade Federal do Esp\'irito Santo, 29075-910 Vit\'oria, ES, Brasil}

\affiliation[d]{Instituto de F\'{\i}sica, Universidade Federal da Bahia, 40210-340 Salvador, BA, Brasil}

\affiliation[e]{Instituto Federal de Educa\c c\~ao, Ci\^encia e Tecnologia da Bahia, 40301-015, Salvador, BA, Brazil}

\affiliation[f]{Observat\'orio Nacional, 20921-400 Rio de Janeiro, RJ, Brasil}

\abstract{The current observational tensions in the standard cosmological model have reinforced the research on dynamical dark energy, in particular on models with non-gravitational interaction between the dark components. \MB{Analyses of} late-time observables like type Ia supernovas (SNe Ia) and large-scale structures (LSS) \MB{are not conclusive about the presence of energy flux between dark energy and dark matter}, while the anisotropy spectrum of the cosmic microwave background (CMB) \MB{is fully consistent} with no interaction at all. As background and visible matter tests are \MB{less sensitive} to the suppression/enhancement in the dark matter power spectrum, which is a characteristic of interacting models, while the CMB spectrum is strongly affected by it, this could be the origin of those results. In order to confirm it and at the same time to rule out the role of possible systematics between early and late-time observations, the use of a low redshift observable sensitive to the gravitational potential generated by dark matter is crucial. In the present paper, we investigate the observational viability of a class of interacting dark energy models, namely with energy exchange between vacuum-type and dust components, in the light of the Dark Energy Survey (DES) observations of galaxy weak lensing, in the context of a spatially-flat Friedmann-Lemaître-Robertson-Walker spacetime. The best fit of our analysis is \MB{entirely consistent} with null interaction, confirming the CMB based constraints.}

\begin{document}
\maketitle

\section{Introduction}
\label{Introduction} 

{{As observational precision increases, some tensions between different data sets have emerged in the context of the standard $\Lambda$CDM model \cite{DiValentino:2021izs,micol,Benetti:2021div}, notably between the expansion rates derived from local measurements of type Ia Supernovae (SNe Ia) \cite{Riess:2019cxk} and from the temperature anisotropies in the Cosmic Microwave Background (CMB) \cite{Aghanim:2018eyx}, as well as between the estimates of the matter fluctuation amplitude from CMB and cosmic shear data \cite{Hildebrandt:2018yau, MacCrann:2014wfa}.}} These issues have reinforced the research on alternative models, such as those with dynamical dark sector or modified gravity theories \cite{Graef:2018fzu, Benetti:2017juy,
Handley:2019tkm, DiValentino:2019qzk, Bernal:2016gxb, Guo:2018ans, Vattis:2019efj, Capozziello:2020nyq, Benetti:2020hxp, Benetti:2019gmo, Pan:2019gop,Poulin:2018cxd,Knox:2019rjx,Yang:2021hxg,Alcaniz:2019kah, Keeley:2019esp, Li:2019yem, Li:2020ybr, Szydlowski:2017wlv,Li:2019san, Szydlowski:2018kbk,Shafieloo:2016bpk}. 

In the case of dynamical models, an important aspect is how to treat the dark energy (DE) perturbations.  
A possible procedure is to decompose the dynamical dark energy into a pressureless, clustering component and a vacuum-type term with an Equation-of-State (EoS) parameter $w = -1$ \cite{Zimdahl:2005ir,wands2,Borges:2013bya,Alcaniz:2012mh,Gorini:2007ta}. In this case, it is possible to show \cite{micol} that the latter does not cluster in the limit of sub-horizon scales, while the former can be reinterpreted as dark matter. In this framework, a flux of energy between the dark components generally occurs, whose observational signature would indicate a dynamical nature of the original DE field. Such an energy flux violates adiabaticity and characterises the so-called interacting DE models (iDE).

In a previous work \cite{micol}, it was shown that a joint analysis of CMB (Planck 2015) \MB{lensed} data and SNe Ia observations \MB{was unable to definitely constrain} the interaction parameter $\alpha$ of a particular class of iDE models, 
\MB{with confidence intervals consistent with zero at $1\sigma$ level,} corroborating results of similar studies \cite{wands2,Aurich:2017lck}. The dark sector was modeled as a non-adiabatic generalised Chaplygin gas, decomposed into energy-exchanging vacuum-type and dust components. This analysis was updated with the Planck 2018 likelihoods \cite{Aghanim:2018eyx}, where a non-zero spatial curvature has also been taken into account \cite{Benetti:2021div}.

Observational tests of interacting models should, however, take into account the fact that, with the above decomposition, only dark matter interacts with dark energy, while the baryonic content is conserved and well constrained by CMB observations and the primordial abundance of light elements. At the background level, this subtlety is not important and has no effect on tests like the determination of distances with standard candles like SNe Ia or standard rules like {baryon acoustic oscillations} (BAO) or the CMB acoustic scale. Nevertheless, as we will review here, at the perturbation level the interaction between the dark components leads to a suppression (enhancement) in the dark matter power spectrum due to the homogeneous creation (annihilation) of dark matter. As visible matter is formed by baryons, this is \MB{less} relevant for the observation of the galaxy power spectrum. It also affects indirectly the baryonic power spectrum, but the effect \MB{could be small enough} to be incorporated into the galaxy bias, \MB{depending on the precision of the galaxy survey and the strength of the interaction}. On the other hand, it is very significant when fitting the full CMB anisotropy spectrum, because the latter is sensitive to the gravitational potential, which -- as dark energy does not cluster -- is determined by the density contrasts of both dark matter and baryonic matter. This is the primary reason for interactions in the dark sector to be so strongly constrained by CMB observations in comparison to background or {large-scale structures} (LSS) tests, which seemed to favor the presence of energy flux from dark energy to dark matter \cite{Alcaniz:2012mh,non-adiabatic2,cassio}. Recent observations of LSS by {Dark Energy Spectroscopic Instrument} (DESI) also suggest dynamical dark energy \cite{desi}, although it depends on the paremetrisation adopted for the the dark energy equation of state.

\MB{We should, however, be aware of the existence of the} above mentioned tensions between the early-time observations involving CMB and late-time tests using SNe Ia and weak lensing, namely between the {Hubble constant}, $H_0$, and {the amplitude of matter perturbations at $8$Mpc}, $\sigma_8$, values derived from those observables. If such tensions are due to unknown systematics, it would be possible, in principle, to conciliate CMB, LSS and background constraints on the interaction parameter. In this scenario, the test of interacting models against weak lensing observations, as performed here, may be crucial to rule out such a possibility, because galaxy weak lensing is a low-$z$ observable that, like the CMB spectrum, is however sensitive to the gravitational wells produced by both dark and baryonic matter.

The paper is structured as follows. In the next section, we present the parametrisation adopted for the interaction, with the dark sector modeled as a non-adiabatic generalised Chaplygin gas split into interacting vacuum-type and pressureless components. The corresponding perturbation equations are also presented, as well as the dark matter and baryonic power spectra, showing the characteristic suppression/enhancement in the former. In Section 3 we focus on the spectrum of convergence, comparing interacting cases with the standard one. In Section 4 we present our results on the shear spectra, in particular the best fit for the interaction parameter and the main cosmological parameters on the light of the \MB{DES Y1} data. Section 5 closes the paper with some final remarks.

\section{Interacting model}
\label{Sec:Theory}

\subsection{Background}

In a spatially flat {Friedmann-Lemaître-Robertson-Walker} (FLRW) universe formed by baryons {(b)} and by a pressureless dark matter fluid {(dm)} interacting with a vacuum component {($\Lambda$)}, the Friedmann and energy balance equations are written as\footnote{We will use $8\pi G = c = 1$.}
\begin{eqnarray} \label{Friedmann}
3H^2 = \rho_m + \Lambda,\\ \label{conservation}
\dot{\rho}_{dm} + 3H\rho_{dm} = \tilde{\Gamma} \rho_{dm} = -\dot{\Lambda},\\ \label{baryons}
\dot{\rho}_b + 3H\rho_b = 0,
\end{eqnarray}
where $\rho_m = \rho_b + \rho_{dm}$ is the total matter density, and the function $\tilde{\Gamma}$ is the rate of dark matter creation (negative in the case of annihilation). By adding (\ref{conservation}) and (\ref{baryons}) we obtain the total matter balance equation
\begin{equation} \label{balance}
    \dot{\rho}_{m} + 3H\rho_{m} = {\Gamma} \rho_{m} = -\dot{\Lambda},
\end{equation}
with $\Gamma = \tilde{\Gamma}\, (\rho_{dm}/\rho_m)$.

We consider a vacuum term evolving as
\begin{equation} \label{Lambda}
\Lambda = \sigma H^{-2\alpha},
\end{equation}
where the interaction parameter $\alpha > -1$ and $\sigma = 3 (1 - \Omega_{m}) H_0^{2(\alpha+1)}$, $\Omega_m$ being the present matter density parameter. \MB{This power-law parametrisation is general enough to test the presence of interaction and includes three natural particular cases: the standard model with $\alpha = 0$; a matter constant creation rate with $\alpha = -1/2$; and a creation rate proportional to the expansion rate $H$, with $\alpha = -1$. Note that the constant $\sigma$ was chosen so that, for $z = 0$, we have $\Lambda = 3H_0^2 (1 - \Omega_m)$.} From the above equations, we can show that
\begin{equation} \label{Gamma}
\Gamma = -\alpha \sigma H^{-(2\alpha +1)},
\end{equation}
and the {background evolution} is given by
\begin{equation}\label{eq:E}
E(z) = {H(z)}/{H_0} = \sqrt{\left[ (1-\Omega_{m}) + \Omega_{m} a^{-3(1+\alpha)} \right]^{\frac{1}{(1+\alpha)}}},
\end{equation}
where the radiation component was neglected, {and $a$ is the scale factor.}

Eq.~(\ref{eq:E}) is the Hubble function of a generalised Chaplygin gas (GCG) \cite{cg1,Dev:2002qa,Alcaniz:2002yt,cg2,cg3}, which behaves like cold matter at early times and as a cosmological constant in the asymptotic future. More precisely, it is a non-adiabatic Chaplygin  gas \cite{Borges:2013bya,Alcaniz:2012mh,non-adiabatic2,wands2,bb}, because the vacuum component does not cluster and hence no pressure term appears in the perturbation equations. Consequently, the power spectrum does not present oscillations and instabilities characteristic of the adiabatic version. Conserved baryons are included in the term proportional to $(1+z)^3$ in the binomial expansion of (\ref{eq:E}). The standard model is recovered for $\alpha = 0$.

\subsection{Perturbations}

The perturbation equations for baryons remain the same as in the $\Lambda$CDM model. Assuming no momentum transfer between the dark components\footnote{For models with momentum transfer, see e.g. \cite{Beltran}.}, the Poisson and dark matter perturbation equations in the longitudinal gauge assume the form \cite{micol,Benetti:2021div,Salzano:2021zxk}
\begin{equation}\label{thetad2}
\theta'_{dm}+\mathcal{H}\theta_{dm}-k^2\Phi=0,
\end{equation}
\begin{equation}\label{deltad2}
\delta'_{dm}-3\Phi'+\theta_{dm}=-\frac{aQ}{\rho_{dm}} \left[ \delta_{dm} - \frac{1}{k^2} \left( k^2 \Phi + \frac{Q'}{Q}\theta_{dm} \right) \right],
\end{equation}
\begin{equation} \label{Poisson}
-k^2 \Phi=\frac{a^2}{2} (\rho_{dm} \delta_{dm} + \rho_b \delta_b) - \left( \frac{a^3 Q}{2} - \frac{3a^2}{2}\mathcal{H}\rho_m \right)\frac{\theta_{dm}}{k^2},
\end{equation}
where $\mathcal{H} = aH$, $Q = \Gamma \rho_m = - \dot{\Lambda}$, the prime means derivative w.r.t. the conformal time, $\theta_{dm}$ is the dark matter velocity potential, $\Phi$ is the gravitational potential, {and $k$ is the comoving wavenumber.}

In the sub-horizon limit they are reduced to
\begin{equation}\label{ol}
\theta'_{dm}+\mathcal{H}\theta_{dm}-k^2\Phi=0,
\end{equation}
\begin{equation}\label{ol1}
\delta'_{dm}+\theta_{dm}=-\frac{aQ}{\rho_{dm}} \delta_{dm},
\end{equation}
\begin{equation}\label{ol2}
-k^2 \Phi=\frac{a^2}{2} (\rho_{dm} \delta_{dm} + \rho_b \delta_b).
\end{equation}
For the vacuum term, we have $\delta \Lambda = -aQ\theta_{dm}/k^2$ and $\delta Q = Q'\theta_{dm}/k^2$, which shows that the vacuum perturbations are negligible for sub-horizon modes.

As the vacuum component does not cluster in sub-horizon scales, there is no entropic sound speed, i.e.
$c_s^2\propto \delta\rho_{\Lambda}=0$. In fact, a scale-independent second-order differential equation for total matter can be derived by combining equations ($\ref{ol}$)-($\ref{ol2}$), 
\begin{equation} \label{delta_m}
\delta_m^{''}+\bigg(\mathcal H+\frac{aQ}{\rho_m}\bigg)\delta_m^{'}+\bigg[\bigg(\frac{aQ}{\rho_m}\bigg)'+\frac{aQ}{\rho_m}\mathcal H-\frac{1}{2}a^2\rho_m\bigg]\delta_m=0.
\end{equation}
Still neglecting the coupling with radiation, for conserved baryons we have a similar equation with $Q=0$,
\begin{equation} \label{delta_b}
\delta_b^{''}+\mathcal H \delta_b^{'}-\frac{1}{2}a^2\rho_m\delta_m=0.
\end{equation}

\begin{figure}[t]
\centerline{\includegraphics[scale=0.32]{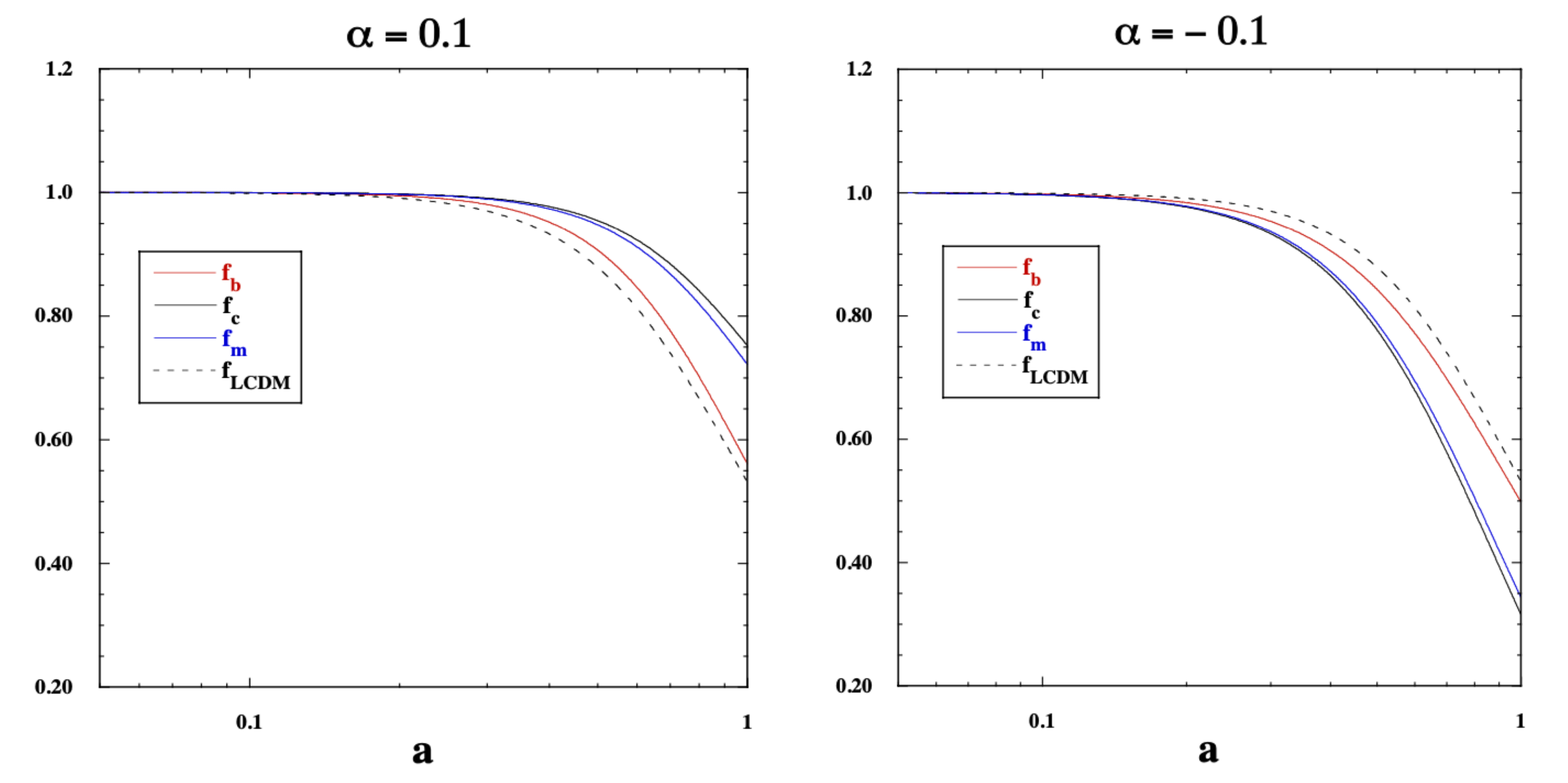}} 
\caption{Evolution of the growth rates $f_i = d\ln \delta_i/d\ln a$ for total matter ($f_m$), cold dark matter ($f_c$) and baryonic matter ($f_b$), for positive and negative values of the interaction parameter. The $\Lambda$CDM growth rate is also shown for comparison. The density parameters $\Omega_b = 0.05$ and $\Omega_m = 0.32$ were used for baryons and total matter, respectively \cite{Salzano:2021zxk}.}
\label{Fig.1}
\end{figure}

The interaction terms in (\ref{delta_m}), proportional to $Q$, cause a suppression (or enhancement) in the total matter density contrast (as well as in the dark matter contrast, of course), which makes observables like CMB and weak lensing good probes of interaction, since they are sensitive to the clustering of total matter. The baryonic contrast is also indirectly affected through the last term of (\ref{delta_b}), but, \MB{if the interaction is weak}, the effect \MB{could be} small enough to be incorporated into a galaxy bias, as shown in Fig.~\ref{Fig.1}, extracted from \cite{Salzano:2021zxk}. 


\section{Convergence spectra}

\MB{The convergence power spectrum is generally written as} \cite{Bartelmann,Sato:2013mq,Troxel,Meneghetti}
\begin{eqnarray}
    C^{\kappa}_l = \frac{\sigma_8^2}{4} \int_0^{\infty} \left[ \rho_m a^2 D(a)  q(\chi) \right]^2 {\cal P} \left( \frac{l}{\chi} \right) d\chi,
\end{eqnarray}
\MB{which evidences the role of the gravitational potential via the Poisson equation (\ref{ol2}}),
\begin{equation}
    \nabla^2 \Psi = \frac{1}{2} \rho_m a^2 \delta_m,
\end{equation}
\MB{where the high order corrections in (\ref{Poisson}) are neglected at sub-horizon scales. In the expression above, $\chi$ is the proper radial distance, and $D(a)$ is the normalised growing function, in such a way that the matter power spectrum for $k = l/\chi$ is given by}
\begin{equation}
    {\cal P}(a,k) = \sigma_8^2 D^2(a) {\cal P}(k).
\end{equation}
The lens efficiency $q$ is defined as
\begin{equation}
    q(\chi) = \int_{\chi}^{\infty} d\chi' n(\chi') \left( \frac{\chi'-\chi}{\chi'} \right),
\end{equation}
where $n(\chi)$ is the normalised galaxy density distribution\footnote{In next section we adopt the CLASS default for $n(a)$.}.

\MB{In the standard case of a $\Lambda$CDM model, the background matter density scales as}
\begin{equation}
    \rho_m = 3H_0^2 \Omega_m a^{-3},
\end{equation}
\MB{leading to the convergence spectrum}
\begin{equation} \label{convergence}
    C^{\kappa}_l = \frac{9}{4} H_0^4 \Omega_{m}^2 \sigma_8^2 \int_0^{\infty} \left[ \frac{q(\chi) D(a)}{a(\chi)} \right]^2 {\cal P} \left( \frac{l}{\chi} \right) d\chi.
\end{equation}
In the case of the interacting model, using (\ref{Lambda}) and (\ref{eq:E}) we obtain the matter density
\begin{equation}
     \rho_m = 3H^2 - \Lambda = 3H_0^2 \Omega_m a^{-3} \beta(a),
\end{equation}
where
\begin{equation}
\beta(a) = \frac{a^3}{\Omega_m} \left[ E(a)^2-(1-\Omega_m) E(a)^{-2\alpha} \right],
\end{equation}
which reduces to $\beta = 1$ for $\alpha = 0$. Therefore, the convergence spectrum can be obtained by just inserting the factor $\beta(a)$ in the integral (\ref{convergence}),
\begin{equation} \label{convergence2}
    C^{\kappa}_l = \frac{9}{4} H_0^4 \Omega_{m}^2 \sigma_8^2 \int_0^{\infty} \left[ \frac{q(\chi) D(a) \beta(a)}{a(\chi)} \right]^2 {\cal P} \left( \frac{l}{\chi} \right) d\chi.
\end{equation}

\begin{figure}[t]
\centerline{\includegraphics[scale=0.6]{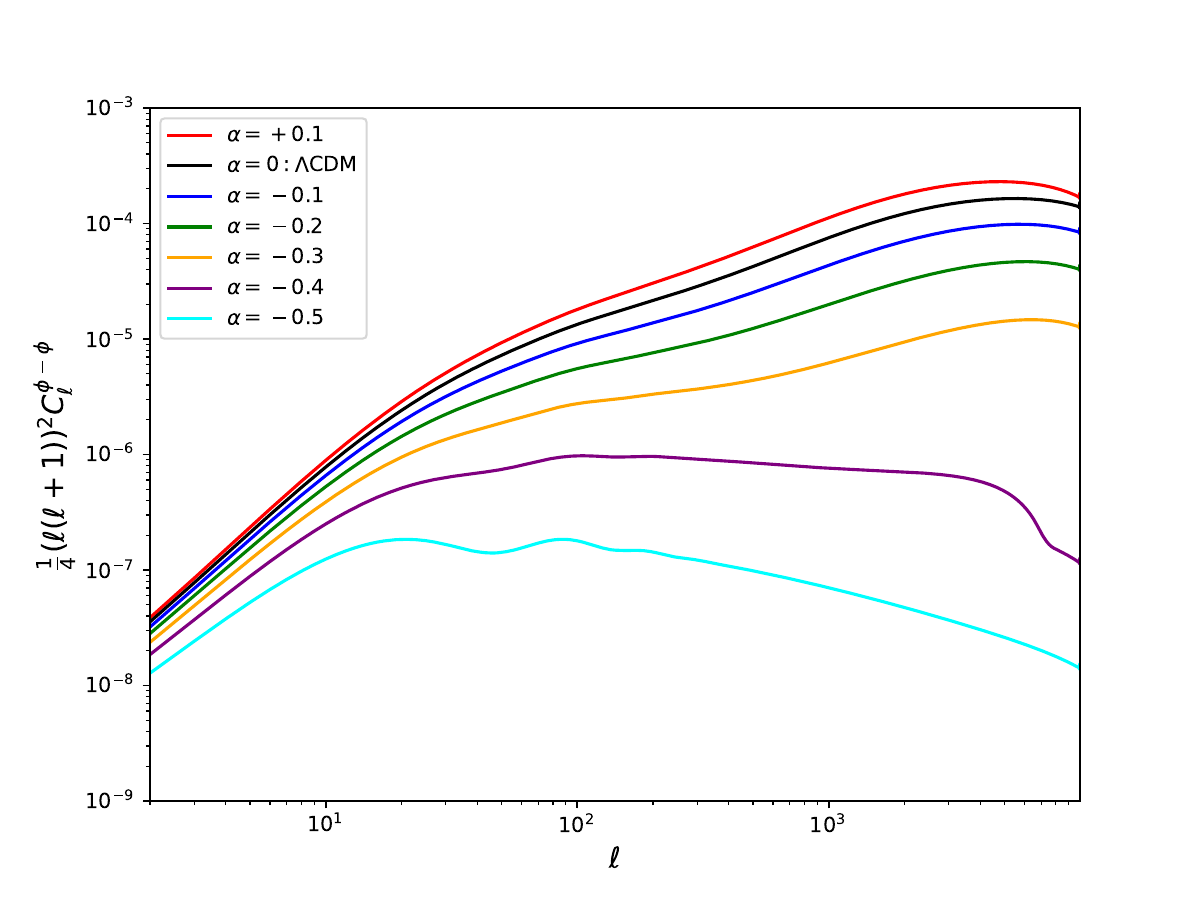}} 
\caption{Non-linear galaxy lensing potencial power spectra for different values of the interacting parameter. For $\alpha=0$ the $\Lambda$CDM model is recovered.}
\label{Fig.2}
\end{figure}

We show in Fig.~\ref{Fig.2} the resulting spectra for different values of the interaction parameter, including the $\Lambda$CDM particular case with $\alpha = 0$. As expected from our previous discussions, for negative $\alpha$, i.e. an energy flux from dark energy to dark matter, there is a suppression in the spectra, caused by the suppression in the total matter contrast that sources the gravitational potential, as already shown in Fig.~\ref{Fig.1}. The opposite occurs for positive $\alpha$. 
We have generated these spectra using the Botzmann solver code \textit{Cosmic Linear Anisotropy Solving System} (CLASS) assuming a non-linear ${\cal P}(k)$, using the fiducial halofit provided by CLASS, which, strictly speaking, is only valid for the $\Lambda$CDM model. Neverthless, this is not important for small values of $\alpha$ and, in any case, the suppression/enhancement is already noticeable at linear scales. 

\section{Results and discussion}
\label{Sec:Analysis}

As ellipticities are easier to measure than magnification, the convergence spectrum is not directly tested, but, instead, the related shear spectra \cite{Bartelmann,Meneghetti}. The same suppression or enhancement \MB{present in} the former will also affect the latter, turning shear data a good probe of dark sector interactions. In this section we test the above interaction parametrisation against the first year shear data of the Dark Energy Survey collaboration (DES Y1) \cite{DESY1}. The complete data include galaxy clustering, galaxy-galaxy shear and cosmic shear. As discussed in the Introduction and explicitly shown in Section~2, the baryonic power spectrum is \MB{only indirectly} affected by the production/annihilation of dark matter and, hence, the first two sets of data are not the most appropriate for our purpose, since any signature of interaction could be absorbed in the galaxy bias \MB{if the interaction is not very strong}. Therefore, we will first focus on the fit of the cosmic shear DES Y1 likelihood, for which the effect would be more pronounced. \MB{Let us stress that the results from the cosmological parameter constraints based on the shared data from both the first (DESY1) and third (DESY3) releases of the DES project are in complete agreement, with no significant deviations or improvements in precision \cite{Kilo-DegreeSurvey:2023gfr,Garcia-Garcia:2024gzy}. Therefore, using the first release offers valuable insights into the parameter constraints of this data survey, without any loss of generality for our results.}

\begin{figure}[t]
\centerline{\includegraphics[scale=0.24]{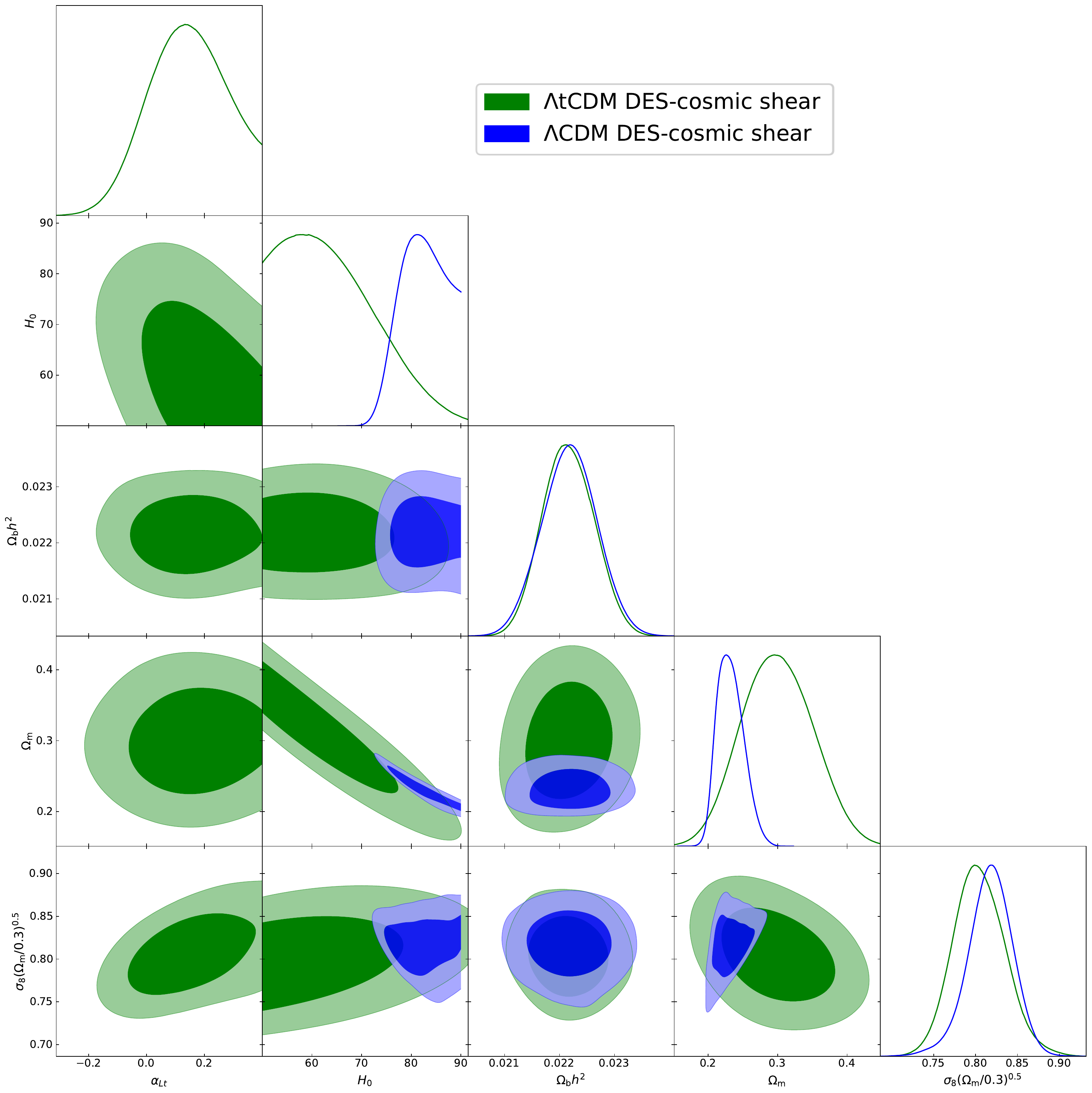}} 
\caption{Confidence regions for \MB{DES Y1} cosmic shear data \cite{DESY1} with the interacting (red) and $\Lambda$CDM (blue) models. A prior on the baryon density parameter was used \cite{Cooke:2017cwo}.
}
\label{Fig.3}
\end{figure}

The analysis was performed using the \MB{freely available} Boltzmann solver's CLASS code \cite{CLASS},
adapted to the interaction model with \MB{the free parameter} $\alpha$, and the \MB{sampler package} COBAYA \cite{Cobaya}.
\MB{We leave the interaction parameter and the cosmological quantities $H_0$, $\Omega_b$ and $\Omega_m$ free to vary, with flat priors around the standard model best-fit values. Also, a Gaussian prior $\Omega_b h^2 = 0.02237 \pm 0.00070$ was imposed on the baryon density from primordial deuterium abundance \cite{Cooke:2017cwo}. Our dataset consists in the DES Y1 cosmic shear likelihood, but we also explore for a comparison the joint best-fit of the cosmic shear, galaxy-galaxy shear and galaxy clustering likelihoods of DES \cite{DESY1} as well as the CMB data from Planck2018 \cite{Planck:2019nip}, i.e. TTTEEE+lowE+lensing data.} 
Our results are shown in Fig.~\ref{Fig.3}, i.e. the $2$-dimensional confidence contours and posteriors function for each free parameter, and also the clustering parameter $\sigma_8$. For comparison, we present the corresponding confidence regions for the $\Lambda$CDM model as well, using the DES cosmic shear. The results generally confirm those obtained with Planck's data \cite{micol,Benetti:2021div}, in particular for the interaction parameter, for which we have a slightly positive value (meaning energy transfer from dark matter to dark energy), but quite compatible with no interaction. The best-fits for the matter density and $\sigma_8$ are also in good agreement with the standard values derived from CMB observations. An important exception is given by the Hubble parameter, whose best-fit is smaller than the CMB standard value, due to an anti-correlation in the plane $\alpha \times H_0$, opposite to the positive correlation found from the CMB \MB{lensed} data \cite{micol,Benetti:2021div}. The $1\sigma$ confidence region for $H_0$ is, however, fully compatible with the standard Plancks's 2018 best-fit, although in marginal tension with supernova-based measurements of the Hubble parameter. 

\begin{figure}[t]
\centerline{\includegraphics[scale=0.33]{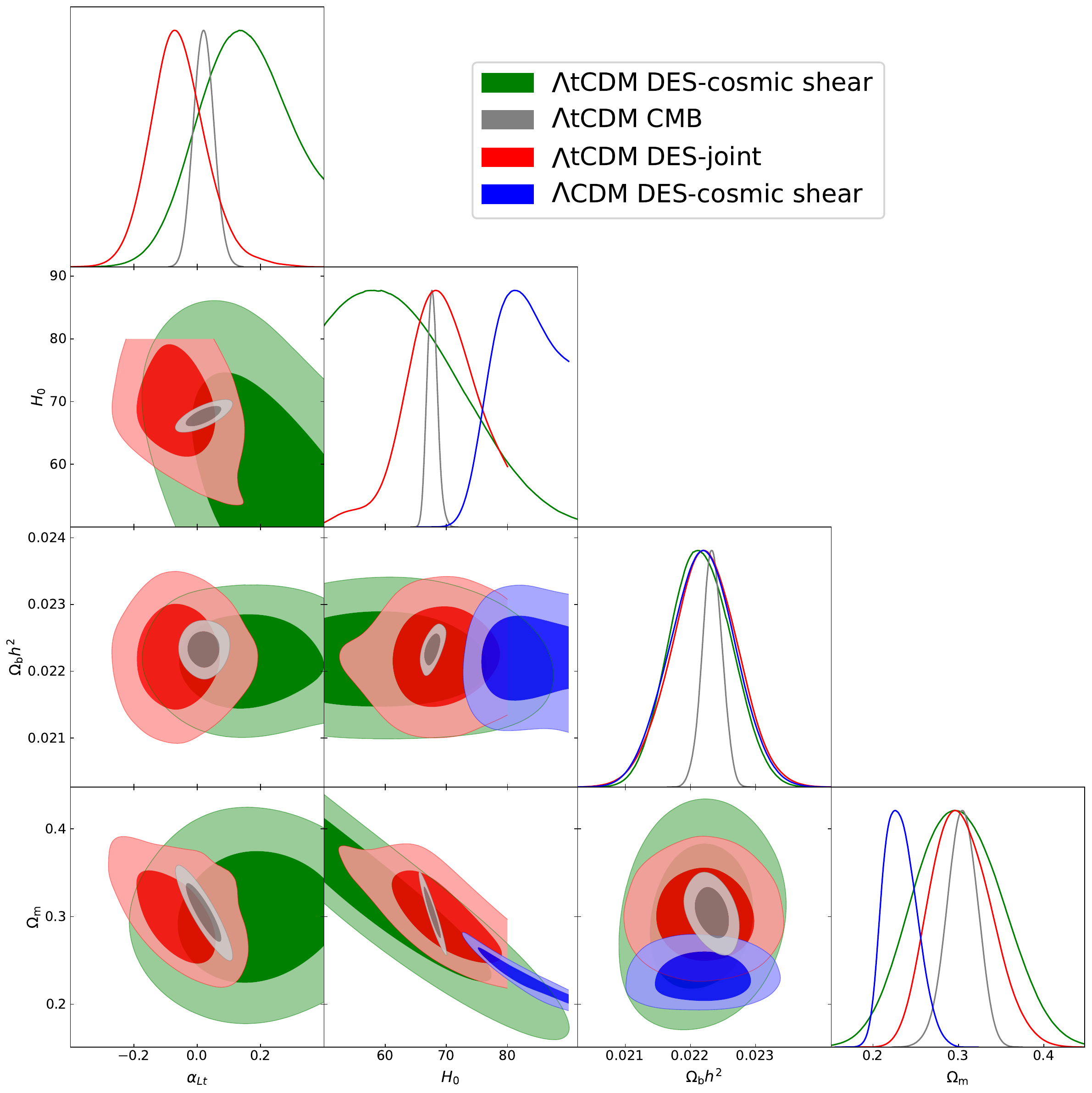}} 
\caption{Confidence regions for \MB{DES Y1} cosmic shear data \cite{DESY1} for the interacting (green) and $\Lambda$CDM (blue) models. We also show the joint analysis of cosmic shear + galaxy clustering + galaxy-galaxy shear for the interacting model (red), as well as the confidence contours for CMB temperature anisotropy \MB{lensed} data (gray). A prior on the baryon density parameter was used \cite{Cooke:2017cwo}. 
}
\label{Fig.4}
\end{figure}

In Fig.~\ref{Fig.4} we superpose the previous analysis with the interacting model joint data of DESY1 (red contours), as well as the confidence regions derived from the \MB{lensed} Planck's 2018 CMB temperature anisotropies \cite{Aghanim:2018eyx} (gray contours). As we are using the suppressed/enhanced total matter to fit the galaxy contrast, which, as discussed above, is weakly affected by the interaction, the joint analysis puts stronger constrains on the interaction parameter. On the other hand, we note the above mentioned orthogonality between the CMB and cosmic shear confidence regions in the $\alpha \times H_0$ plane, with the intersection of $1\sigma$ contours almost ruling out any signature of dark sector interaction.
We also report the mean values at $68\%$ confidence level from the interaction model analyses using \MB{DES Y1} data in Table \ref{tab:results}.

\begin{table}[h!]
    \centering
    \begin{tabular}{|c|c|c|}
        \hline
        \textbf{} & \textbf{DES shear} & \textbf{DES joint} \\
        \hline
        $\alpha$ & $0.15 \pm 0.13$ & $-0.06 \pm 0.09$ \\
        $H_0$ & $64.03 \pm 8.45$ & $68.77 \pm 5.56$ \\
        $\Omega_\mathrm{b} h^2$ & $0.0222 \pm 0.0005$ & $0.0222 \pm 0.0005$ \\
        $\Omega_\mathrm{m}$ & $0.301 \pm 0.049$ & $0.303 \pm 0.034$ \\
        $\sigma_8$ & $0.816 \pm 0.085$ & $0.762 \pm 0.055$ \\
        \hline
    \end{tabular}
    \caption{Mean values for the interacting model using \MB{DES Y1} cosmic shear data and joint analysis of cosmic shear + galaxy clustering + galaxy-
galaxy shear.}
    \label{tab:results}
\end{table}

\section{Final Remarks}
\label{Sec:Conclusions}

Models with non-gravitational interaction between the dark components has been exhaustively studied from the theoretical and observational point of views, with the use of diverse parametrisations for the dark energy-dark matter coupling. From the theoretical viewpoint, several motivations have been pointed for justifying the presence of interaction, as for example an unified conception of the dark sector, a possible dynamical nature of dark energy or the creation of particles from the vacuum by the cosmic expansion. From the observational viewpoint, conflicting results have been derived for the sign of the interaction, i.e. whether the flux energy is from dark energy to dark matter or the reverse, depending on the parametrisation adopted for the interaction and the particular probes used in the observational tests, \MB{especially those} involving background and large scale structure data. Actually, no definite conclusion about the presence of interaction has been taken from these tests at all. On the other hand, the hope of solving or alleviating the current tensions in cosmology has not been confirmed either, because of the opposite correlations found between the interaction parameter and the Hubble and $\sigma_8$ parameters.

Nevertheless, we arrive at a clearer picture when testing such models against the spectrum of anisotropies of the cosmic microwave background. In this case, the presence of interaction is tightly constrained and the standard $\Lambda$CDM model favoured. The essential reason is that the CMB anisotropies directly probe the gravitational wells formed by dark matter and, therefore, are very sensitive to the suppression or enhancement in the dark matter clustering characteristic of interacting models. Although it is generally possible to compensate this effect and fit the first acoustic peak of the CMB spectrum by adjusting the Hubble, curvature and matter density parameters, the secondary peaks and the spectrum tail become badly fitted, \MB{especially in view} of the tight constraints imposed on the baryon density by primordial nucleosynthesis.

This is not, however, a conclusive picture at least until the current tensions between early and late times cosmological probes are resolved. The presence of unknown systematics in the CMB data or new physics at early times could in principle alleviate the constraints the former impose on interacting models. A way of verifying or ruling out this possibility is to consider a low redshift probe that would also be sensitive to the suppression/enhancement of gravitational wells due to homogeneous creation/annihilation of dark matter. A good candidate is galaxy weak lensing, and in this paper we used the \MB{DES Y1} cosmic shear data to test a particular interaction parametrisation, modeled by a uni-parametric, non-adiabatic generalised Chaplygin gas. Although the data still do not present the precision reached by other probes, our best-fit corroborates the CMB \MB{lensed} data fit in favour of null interaction, a tendency that is reinforced when both probes are superposed, owing to the orthogonality between their confidence regions in the $\alpha \times H_0$ plane. This result could be confirmed when cosmic shear data become more precise and reliable, definitely ruling out the interaction hypothesis, at least in the context of a Chaplygin gas unified model of the dark sector.

\section*{Acknowledgements}

We are thankful to H. A. Borges, R. von Marttens and F. N. Toscano for useful discussions and suggestions. MB thanks support of the Istituto Nazionale di Fisica Nucleare (INFN), sezione di Napoli, iniziativa specifiche QGSKY. PTS was supported with a PhD grant from CAPES (Brazil). SC acknowledge support from CNPq (Brazil). JCF thanks support of CNPq and FAPES (Brazil). The authors thank the use of CLASS and COBAYA codes.

\end{document}